\documentstyle[11pt,newpasp,twoside,epsf]{article} 
\def\edcomment#1{\iffalse\marginpar{\raggedright\sl#1\/}\else\relax\fi}
\reversemarginpar

\begin{document}
\title{The fine spatial structure of methanol masers 
as an evidence in support of their connection with bipolar outflows
}
\author{V.I. Slysh, I.E. Val'tts and S.V. Kalenskii}
\affil{Astro Space Center, Lebedev Physical Institute,
Profsoyuznaya str. 84/32, 117 810 Moscow, Russia}

\begin{abstract}
We studied class I methanol masers in the transition $7_0-6_1A^+$
at the frequency 44 GHz with the VLA. The 
observations on the VLA were made with the angular resolution 
0$''$.1, which was the highest at 
the moment. It was shown that the masers consist of chains of 
unresolved spots located on 
curved lines or arcs. The length of such arcs is from 20 to 1000 AU 
and the brightness 
temperature of the strongest masers exceeds 3.6x10$^8$ K. The observed 
location of maser spots 
is in agreement with their position at the border line between 
molecular outflows and 
surrounding molecular clouds. The high brightness temperature 
implies that the maser 
condensations have enhanced abundance of methanol due to evaporation 
of methanol from 
the surface of dust grains. The mass of maser condensations is less 
than 4x10$^{-5}$~M$\odot$ and 
corresponds to planetary masses. 
\end{abstract}

\section{Introduction}
Class I methanol masers (most of these masers were detected earlier in
our Parkes 44 GHz survey - Slysh et al. 1994)
have been studied at 44 GHz with rather high angular resolution in 
Kogan \& Slysh 1998. Maps of methanol masers with the highest
angular resolution were obtained on the 
BIMA interferometer at frequencies 84 and 95 GHz, with a beamwidth of 
about 5$''$ (Batrla \& Menten 1998, Plambeck \& Menten 1990,
Pratap \& Menten 1992). 
The masers were shown to consist of several unresolved spots, which do not coincide with 
known objects. Now  we report the results of new VLA observations of class I methanol 
masers in the 44-GHz transition, with an angular resolution of 0$''$.1 and studying the 
structures of the  masers based on the VLA maps. 

\section{Observations and data reduction}

Observations of the 44-GHz methanol line were carried out on June 28, 
1998, for five 
sources (M8E, W33Met,  L379IRS3, GGD27 and G14.33-0.64).
The VLA (NRAO, USA) was in the AB configuration. We used eleven of 27 
antennas that were equipped with 7-mm receivers. 
The synthesized beamwidth of the interferometer was from 0$''$.1x/Arikael
0$''$.25 to 
0$''$.1x0$''$.17  
(depending on the direction toward the source). 
The sources were observed with a bandwidth of 3125 kHz, so that the total 
velocity coverage 
was 23.3 km/s.
128 channels were recorded, yielding a velocity resolution of 0.17 km/s. 
M8E and GGD27 
were observed with a higher spectral resolution of 0.042 km/s. Fig.~1
shows representative maps of two masers: M8E and G14.33$-$0.64. The full
description of maser maps is given in the paper of Slysh et al. 1999.

\begin{figure}
\plotone{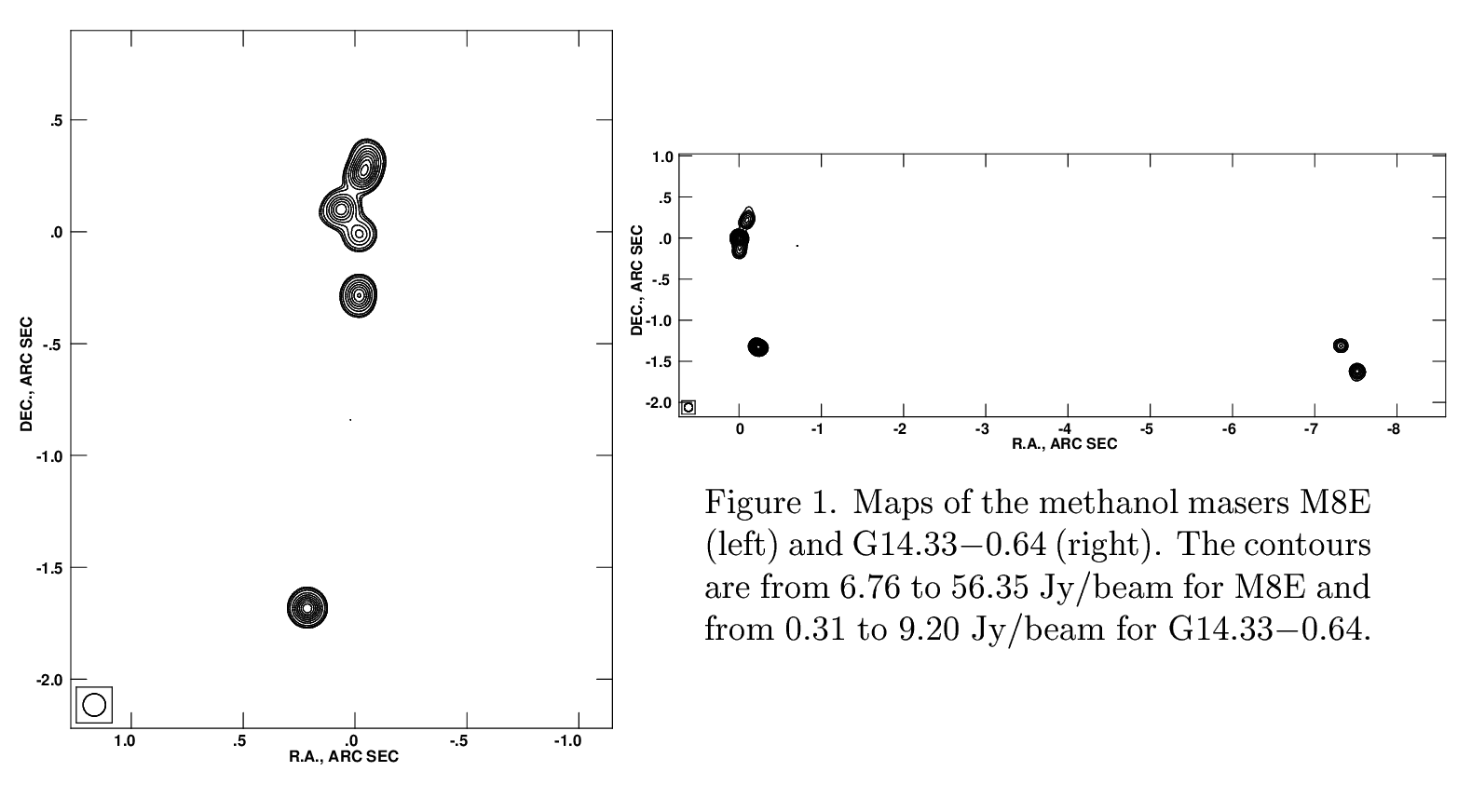}
\end{figure} 

\section{Discussion}

Mapping class I methanol masers with an angular resolution of 0$''$.1
has allowed us to 
reveal their structure. In most cases, the masers form chains or clusters of maser spots. The 
spots themselves are unresolved with the VLA beam, and the
upper limits of the size is typically
0$''$.05. The corresponding lower 
limit on the brightness temperatures for various sources is
between 5.7x10$^6$ and 3.6x10$^8$~K. 
The chains of spots visible in the maps for most of the masers may trace the
geometry of the maser emission regions. One possibility is that the
maser spots are bright 
knots on long filaments (up to 10000 AU) with thickness of less
than 200 AU, which could 
form in regions of intersection of two spherical shock waves, possibly arising during collisions 
of molecular outflows.
In another model in which the chains of maser spots are located on the spherical surface of a 
shock wave, the maser spots are observed at the tangential surface of the shock front, where 
the amplification path is at maximum. The projected sizes of the maser spots can be much 
smaller than their sizes along the line of sight. In this case, the maser spots outline the outer 
contours of outflows.
None of the class I methanol maser spots display the high velocities observed in molecular 
outflows. Their radial velocities are usually at the center of the thermal line, which forms in 
the medium surrounding the molecular cloud, and the velocity dispersion is always smaller 
than the width of the thermal line. If the maser spots are located at the tangential surface of 
a shock wave, they can still move with high velocities, but only in the transverse direction. 
Since the amplification factor in the tangential direction is greater than in the transverse 
direction, the maser emission will be observed with a small line-of-sight velocity. In this 
picture, the maser condensations move with the shock surface, but the only ones that are 
visible are those that move in the transverse direction.
This model can be tested by deriving the tangential velocities from the proper motions of the 
maser spots. The tangential-velocity estimates in W33Met and L379IRS3 that we have 
obtained here are not sufficiently precise to detect proper motions with values
typical for molecular outflows or shock waves. The upper limits on the transverse velocities 
are 30 and 60 km/s for L379IRS3 and W33Met, respectively, which are larger than the 
typical velocities of molecular outflows (20 km/s). More precise proper-motion 
measurements are required to test the propagating shock-wave model for the maser spots.
On the other hand, the maser condensations could be stationary, while the shock flows 
around them. In this picture, the interaction with the shock increases the temperature, 
methanol evaporates from the surfaces of dust grains, and the methanol abundance increases 
in
the condensations to a value sufficient for the generation of maser emission. After the passage 
of the shock, the condensations remain stationary.
The origin of small (100-200 AU) maser condensations is unclear. One possibility is that they 
appear as a result of Rayleigh-Taylor instabilities that arise during the interaction of a shock 
with the surrounding medium. However, we do not have enough observational evidence to 
confirm any specific model. Though the connection of methanol masers with bipolar 
outflows is well established, there are some masers (for example, NGC6334-I(N)) in which 
outflows are not
observed. In addition, in many cases, emission from excited
vibrational levels in the H$_2$
molecule, which is a good tracer of interactions between an outflow and the surrounding 
medium, is not observed. If maser spots are stationary condensations in the surrounding 
medium with a molecular outflow flowing around them, we might expect a correspondence 
between maps of the methanol masers and of the H$_2$ line emission. For the sources we have 
considered here, we found no such correspondence. If the observed projected sizes of the 
maser condensations 100-200 AU are equal to their line-of-sight sizes, we can estimate their 
masses, assuming their density is that of molecular hydrogen,
10$^6$~cm$^{-3}$ (as follows from the 
maser-pumping model (Menten et al. 1998): 8x10$^{28}$~g, or
4x10$^{-5}$~M$\odot$. Such small condensations
would not be detected in thermal molecular lines, and only the maser emission would 
make them visible. The lower limit on the brightness temperature,
3.6x10$^8$~K, corresponds 
to a maser optical depth of the order of -15, and to a methanol column density of the order 
of 10$^{16}$~cm$^{-2}$ (Liechti \& Walmsley 1997). If the linear size of
a maser condensation is 200 AU
and the H$_2$ density is 10$^6$~cm$^{-3}$, the methanol abundance
should be of the order of 3x10$^{-6}$. This is three to four 
orders of magnitude greater than in the interstellar medium. Such a strongly enhanced 
methanol abundance could be a consequence of evaporation of
methanol from the surface 
of dust grains, stimulated by heating from a shock wave. If the maser condensations are 
extended along the line of sight, as in our model where they are localized on the tangential
surface of a shock wave, their line-of-sight sizes may be much greater than their projected 
sizes, and the corresponding estimated masses will be considerably higher.

\section{Conclusions}

We have studied structure of class I methanol masers based on our high-resolution VLA 
maps. A typical pattern formed by these masers is a chain of bright spots located along 
arclike curves with length of several thousands AU. The spots themselves are not resolved 
and have sizes less than several hundred AU. Some sources have several
maser spot groups 
separated by  few tenths of a parsec.
The methanol maser spots have no counterparts in OH 
or H$_2$O masers and do not coincide with radio continuum or infrared sources. At the 
same time, class I methanol masers are tightly connected with molecular outflows. Our results 
support the hypothesis that the methanol masers arise in zones of interaction of molecular 
outflows between themselves or with the surrounding medium.
Proceeding from the fact that the brightness temperatures
of these masers are very high, more 
than 10$^8$~K, we conclude that the maser condensations should have
strongly enhanced 
methanol abundance as a result of methanol evaporation from the surface of dust grains.
The masses of the maser condensations do not exceed
4x10$^{-5}$~M$_\odot$, roughly corresponding to planetary masses.

\acknowledgements The authors are grateful to the VLA
staff for the help with the observations.
The work was done under a partial financial support from the INTAS
(grant No. 97-11451) and the Russian Foundation for
Basic Research (grant No. 01-02-16902).

\end{document}